\documentstyle[12pt]{article}
\def\bcn{\begin{center}}
\def\ecn{\end{center}}
\newcommand{\be}{\begin{equation}}
\newcommand{\ee}{\end{equation}}

\def\barr{\begin{array}}
\def\earr{\end{array}}
\def\and{\qquad {\rm and } \qquad}
\def\etal{ {\it et al.}}
\def\bib{\bibitem}

\def\pb{\: {\rm pb}}
\def\gev{\: {\rm GeV} }
\def\lsim{\:\raisebox{-0.5ex}{$\stackrel{\textstyle<}{\sim}$}\:}
\def\gsim{\:\raisebox{-0.5ex}{$\stackrel{\textstyle>}{\sim}$}\:}
\def\ra{\rightarrow}

\def\ib#1,#2,#3{       {\it ibid.\/ }{\bf #1}, #3 (19#2)}
\def\ap#1,#2,#3{       {\it Ann.~Phys.~(NY)\/ }{\bf #1}, #3 (19#2)}
\def\ijmp#1,#2,#3{     {\it Int.~J.~Mod.~Phys.\/ } {\bf A#1}, #3 (19#2)}
\def\mpl#1,#2,#3 {     {\it Mod.~Phys.~Lett.\/ } {\bf A#1}, #3 (19#2)}
\def\np#1,#2,#3{       {\it Nucl.~Phys.\/ }{\bf B#1}, #3 (19#2)}
\def\npps#1,#2,#3{     {\it Nucl.~Phys.~B (Proc.~Suppl.)\/ }{\bf B#1}
                            , #3 (19#2)}
\def\plb#1,#2,#3{      {\it Phys.~Lett.\/ }{\bf B#1}, #3 (19#2)}
\def\pr#1,#2,#3{       {\it Phys.~Rev.\/ }{\bf #1}, #3 (19#2)}
\def\prd#1,#2,#3{       {\it Phys.~Rev.\/ }{\bf D#1}, #3 (19#2)}
\def\prep#1,#2,#3{     {\it Phys.~Rep.\/ }{\bf #1}, #3 (19#2)}
\def\prl#1,#2,#3{      {\it Phys.~Rev.~Lett.\/ }{\bf #1}, #3 (19#2)}
\def\pro#1,#2,#3{      {\it Prog.~Theor.~Phys.\/ }{\bf #1}, #3 (19#2)}
\def\rmp#1,#2,#3{      {\it Rev.~Mod.~Phys.\/ }{\bf #1}, #3 (19#2)}
\def\sp#1,#2,#3{       {\it Sov.~Phys.-Usp.\/ }{\bf #1}, #3 (19#2)}
\def\zpc#1,#2,#3{      {\it Zeit.~f\"ur Physik\/ }{\bf C#1}, #3 (19#2)}

\begin{document}
\thispagestyle{empty}
\setcounter{page}{0}
\renewcommand{\thefootnote}{\fnsymbol{footnote}}

\begin{flushright}
MPI--PTh/96--74\\[1.7ex]
{\large \tt hep-ph/9608444} \\
\end{flushright}

\vskip 45pt
\begin{center}
{\Large \bf Light Gluino and Tevatron Data}

\vspace{11mm}
{\large Debajyoti Choudhury}\\[1.5ex]

{\sl E-mail:} debchou@mppmu.mpg.de\\[1ex]
{\em Max--Planck--Institut f\"ur Physik,
              Werner--Heisenberg--Institut,\\
              F\"ohringer Ring 6, 80805 M\"unchen,  Germany.}\\
\vspace{50pt}
{\bf ABSTRACT}
\end{center}
\begin{quotation}
A very light gluino ($m_{\tilde g} \lsim 2 \gev$) is still consistent 
with experimental data and is attractive from a theoretical standpoint. 
This has been shown to lead to a small gluino content of the proton. 
We use this effect to demonstrate that such a light gluino could
lead to a striking enhancement, at Tevatron, of monojets 
accompanied by large missing momentum. A reanalysis of the 
existing data may thus rule out a light gluino for a 
common squark mass of upto $\sim 600 \gev$.
\end{quotation}

\newpage
\renewcommand{\thefootnote}{\arabic{footnote}}
As supersymmetry provides one of the best theoretically motivated 
scenarios going beyond the Standard Model (SM), the search for 
superparticles has, understandably, constituted one of the main 
areas of interest in recent experimental endeavours. Negative results 
at both the Tevatron~\cite{Tevatron_bounds} and LEP~\cite{LEP1.5}
have, however, significantly constrained the 
parameter space available to the ($R$-parity conserving) minimal 
supersymmetric standard model (MSSM). Somewhat surprisingly though,
a very light gluino (mass less than a few GeV) may still be 
allowed~\cite{clav92,farr95,barnett}. Since the esistence of such 
a light gluino will drastically alter the signal for supersymmetry, 
considerable effort has been directed towards a close examination
of this scenario, both from a theoretical standpoint~\cite{banks}
as well as a phenomenological one~\cite{farr_pheno}. Additional 
incentive was provided 
by an assertion~\cite{clav92,alphas} that such a particle helps explain 
the apparent discrepancy between the value of $\alpha_s$ determined
by high energy experiments and that expected from an application
of QCD evolution (with the SM quark content) to the same quantity 
measured at low energy experiments. While this claim has been 
contested~\cite{alphas:opp}, like much else associated with this 
scenario, it points to the need of a closer examination of the 
phenomenological consequences associated with the existence of a 
light gluino. In this Letter we shall  undertake this task from the 
point of view of existing Tevatron data on monojets with large 
missing momenta. We demonstrate that a reanalysis of the
existing can lead to either an evidence for a light gluino, or 
in the negative case, to strong constraints on the scenario. 

To appreciate the peculiarities of the scenario, let us, first, 
briefly recapitulate the essential features and the existing 
constraints. ($i$) As the Tevatron limits~\cite{Tevatron_bounds}
on the squark masses no longer 
apply,  the latter need to be consistent only with the lower 
bounds from LEP1. Note, however, that precision electroweak tests 
disfavour $m_{\tilde q} \lsim 60 \gev$ for a light 
gluino~\cite{gg_arc}.
($ii$) Gluino decay: though light, the gluino is not necessarily
the lightest supersymmetric particle (LSP). The lightest 
neutralino (in most cases, the photino) is often the LSP instead 
(and, thus, a viable dark matter candidate~\cite{dark_matter}). 
In such a case, the gluino would decay into a quark-antiquark pair 
and the LSP, the rate depending on the relevant squark mass. 
Negative search results at beam dump experiments~\cite{beam_dump} 
suggests that a light gluino that decays within the detector volume
is ruled out. This would require that the squark mass
be larger than a few hundred GeV~\cite{dawson}. 
However, unlike its heavy 
counterpart, the gluino can now form a relatively stable and light 
bound neutral state~\cite{bound_state,farr95}. The lifetime of the 
bound state is longer than that of the free gluino~\cite{farr_pheno}
and, furthermore, the photino from the decay interacts too weakly with the 
detector to trigger a signal. 
It has been argued though that, for $m_{\tilde g} \gsim 4 \gev$, 
the lifetime is short enough for a missing energy signal to be 
viable and that UA1 data can effectively rule this out~\cite{UA1}. 
What if the gluino is indeed the LSP ? Even then, the photino is 
the lightest colour-singlet supersymmetric particle~\cite{singlet} 
and hence stable. The lightest bound state still decays into the photino 
and all of the above arguments hold.

Constraints on lighter gluinos are obtained from quarkonia decay. 
The window $1.5 \gev \lsim m_{\tilde g} \lsim 4 \gev$ can be ruled 
out~\cite{Upsi} by considering the decay 
$\Upsilon \ra \tilde{\eta} \gamma$ where $\tilde{\eta}$ is the 
pseudoscalar $\tilde{g} \tilde{g}$ bound state. Extending this
analysis to lower masses is difficult as the applicability of 
perturbative QCD becomes questionable~\cite{Upsi}. 
While some constraints on $m_{\tilde g} \lsim 1.5 \gev$ have been 
discussed in the literature, most of these turn out to be weak or 
model dependent. A case in point are the 
constraints~\cite{b_s_gamma} from the $b \ra s \gamma$ process which
depends strongly on squark mixing. The exception are the 
constraints~\cite{4jet_early} deduced from final state correlations
in $e^+ e^- \ra Z \ra 4 jets$. Indeed, a claim~\cite{gouv_mur} has been 
made recently that LEP data~\cite{4jet_opal,4jet_others}
can be used to rule out this window at 90\% C.L. This claim has however
been criticized~\cite{farr_new} 
on the grounds that the jet angular distributions 
are sensitive to, as yet uncalculated, higher order QCD effects. 

\begin{figure}[h]
        \vskip 5in\relax\noindent\hskip -1.8in
              \relax{\includegraphics{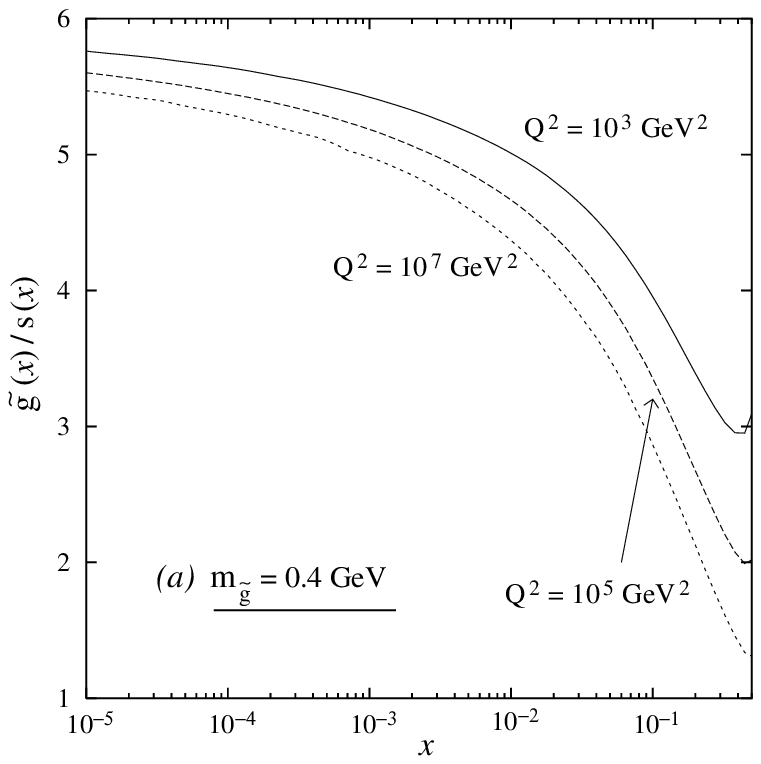}}
        \relax\noindent\hskip 3in
              \relax{\includegraphics{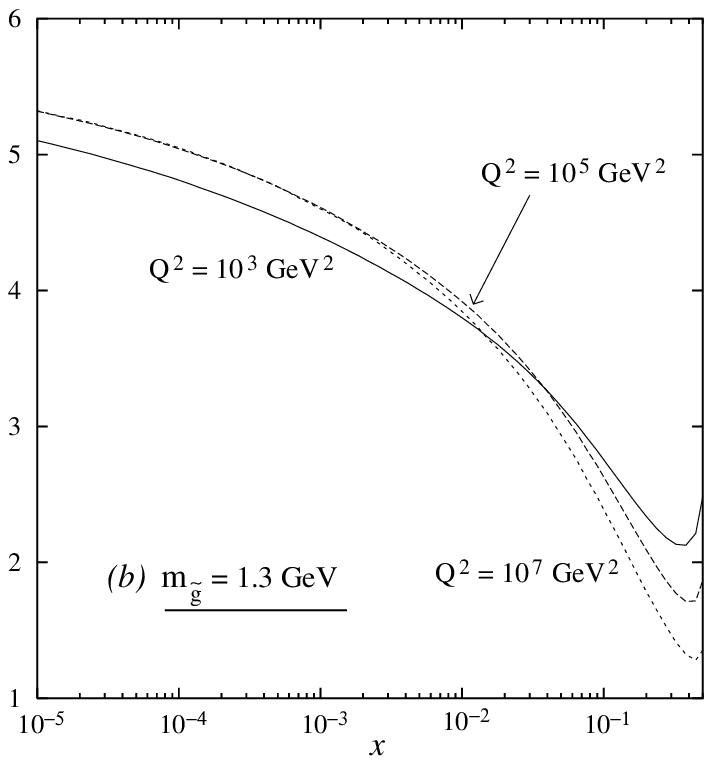}}
        \vspace{-20ex}
 \caption{{\em The ratio of the gluino and strange quark 
          densities~\protect\cite{ruc_vog}
          in the proton as a function of the momentum fraction. 
          {\em (a)} $m_{\tilde g} = 0.4 \gev$, and 
          {\em (b)} $m_{\tilde g} = 1.3 \gev$. 
          }}
 \label{fig:strfn}
\end{figure}

In this Letter, we adopt an approach complementary to that of 
de~Gouv\^ea and Murayama~\cite{gouv_mur}
 and seek to point out that a reanalysis 
of existing Tevatron data can provide important constraints. 
It has been recognized~\cite{ruc_vog,struc2} that the presence 
of light gluinos alters the Altarelli-Parisi evolution of the 
nucleon structure functions in an essential way. 
While it has been argued that this effect is numerically 
too small to be of any relevance to present experiments, we shall 
demonstrate that this is not the case. The important aspect is that 
the proton now has a small but non-zero gluino content. 
In fact, ref.\cite{ruc_vog} (which 
we shall use for the rest of the analysis) explicitly shows that, 
for the mass range in question, the gluino content of the proton 
is roughly 2--5 times that of the strange-quark sea. The ratio 
is only weakly dependent on $Q^2$, but depends significantly on 
the momentum fraction carried by the parton (see Fig.\ref{fig:strfn}). 

This immediately leads to the 
possibility of a {\em resonant} squark production at the Tevatron.
Once produced, the squark would obviously tend to decay into the 
corresponding quark and the gluino. This channel is uninteresting 
though. We rather focus on the suppressed decay channel 
$\tilde{q} \ra q \tilde{\gamma}$. Our process thus is 
\be
        q + \tilde{g} \ra q + \tilde{\gamma} 
           \label{signal}
\ee
This will obviously lead to 
a monojet accompanied by missing transverse momentum (equal to the 
transverse momentum of the jet). The distribution in the latter 
would thus have a peak close to $m_{\tilde q} / 2$ and thus 
could be identified. Although the $s$-channel contribution is the 
dominant one (on account of the resonance), for completeness we 
include the $t$-channel contribution as 
well~\cite{fnote_2}.

The SM background to this process arises from two different 
sources, the straightforward one being $Z + jet$ ($q g \ra q Z$ 
and $ q q \ra g Z$) production with the $Z$ decaying 
invisibly~\cite{fnote_3}. 
Also to be considered are the 
processes $q g \ra q' \tau \nu$ and $q q' \ra g \tau \nu$. Of course,
if the $\tau$ is far away from the jet, it would be recognised as
a thin jet by itself and such configurations can be vetoed. On the 
other hand, if the difference in their azimuthal separation 
($\delta \phi$) and pseudorapidity separation ($\delta \eta$) be 
such that the jet and tau fall within the cone defined by, say,  
$\Delta R \equiv \sqrt{ (\delta \phi)^2 + (\delta \eta)^2} \leq 0.7$,
then these might not be separable and have to be merged to form a 
single jet. 

\begin{figure}[h]
        \vskip 5in\relax\noindent\hskip -0.6in
        \relax{\includegraphics{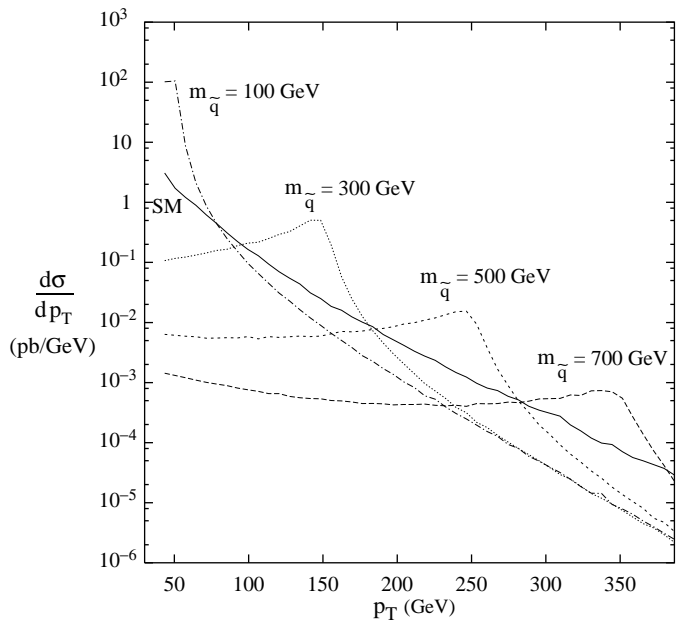}}
        \vspace{-20ex}
 \caption{{\em The $p \bar{p} \ra q + \tilde \gamma + X$ cross-section 
               (at the Tevatron) as a function of the jet $p_T$ for 
               various values of $m_{\rm sq}$. All squarks have been 
               assumed to be degenerate and the cut of 
               eq.(\protect\ref{rapcut}) imposed. Also shown is the 
               SM cross section for (monojet + missing energy). 
               }}
 \label{fig:pT_distrib}
\end{figure}

In Fig.~\ref{fig:pT_distrib}, we show the $p_T$ distribution of 
the process in eq.(\ref{signal}) for various values of squark 
masses. We have assumed here that all squarks are degenerate. 
A minimum of $p_T ({\rm jet})$ of 40~GeV is demanded so that it 
may constitute a clear signal~\cite{fnote_4}. 
To be consistent with detector 
coverage, we have also imposed a cut on jet rapidity:
\be
    | \eta_{jet} | < 3. 
        \label{rapcut}
\ee
We have used here the distributions~\cite{ruc_vog}
for $m_{\tilde g} = 1.3 \gev$ as these lead to the weakest constraints.
To stay on the conservative side, we have further imposed an 
{\em ad hoc} upper bound of $\tilde g (x) / s(x) \leq 3$. The latter, 
though, weakens only the bound on the $\tilde b_{L,R}$ mass.

As expected, the peaks 
lie close to $p_T \sim m_{\tilde q} / 2$. Also shown in the figure is 
the SM background. Thus, with a judicious choice of the $p_T$ window,
a signal to noise ratio larger than unity can be obtained for a wide 
range of squark masses. In Table~\ref{tab:pT_window}
we list the number of events 
in the optimum $p_T$ window for different squark masses. 
\begin{table}[h]
\begin{center}

\begin{tabular}{|r|c|r|r|}
\hline
\multicolumn{1}{|c|} {$m_{\tilde q}$} & 
\multicolumn{1}{c|}  {$p_T$ window} &
\multicolumn{1}{p{0.5in}|} {Gluino} & 
\multicolumn{1}{p{0.45in}|} {SM}
                \\
\multicolumn{1}{|c|} {$(\gev)$} & 
      & 
\multicolumn{1}{p{0.5in}|} {events} & 
\multicolumn{1}{p{0.45in}|} {events} \\
\hline
50 & (40, 65) & 8665 & 5246 \\
100 & (45, 55) & 118004 & 2093 \\
200 & (85, 105) & 8600 & 450 \\
300 & (125, 155) & 1301 & 123 \\
400 & (175, 210) & 190 & 33 \\
500 & (215, 265) & 37.5 & 9.8 \\
600 & (260, 310) & 8.1 & 3.1 \\
700 & (305, 365) & 1.9 & 1.0 \\
\hline
\end{tabular}
\end{center}
   \caption{\em The number of events expected solely from the process of 
     eq.(\ref{signal}) within the $p_T$ window appropriate for a 
     given squark mass. (All squarks are assumed to be degenerate.)
     Also shown are the number of events expected within the SM. 
     An integrated luminosity of $100 \pb^{-1}$ has been assumed.}
   \label{tab:pT_window}
\end{table}
As can be seen from the table, even with an integrated 
luminosity of $100 \pb^{-1}$, 
a significant ${\rm signal} / \sqrt{\rm background}$ 
($S/\sqrt{B}$) ratio can be achieved. 
A strong statement about the existence of such a gluino would, then,
is thus not out of place.

We must, at this stage, point out the potential drawbacks in this 
analysis. As the LSP is mostly photino, its coupling to a quark is 
proportional to the charge of the latter. Furthermore, the quark
content of the proton is dominated by the $u$. Consequently,
the supersymmetric contributions shown in Fig.~\ref{fig:pT_distrib} are 
dominated by the $\tilde u_{L,R}$. Thus one may seek to escape
the bound by postulating the $u$-type squarks to be heavy. 
Such a solution is, however, problematic on more than one count.
For one, apart from introducing an undesirable hierarchy amongst the 
squark masses, a large splitting between isodoublet partners is strongly
disfavoured from considerations of the $\rho$-parameter. Thus, not 
only the $\tilde u_{L,R}$, but the $\tilde d_{L,R}$ will have to be 
heavy. Still, this 
will not solve all problems. 
\begin{table}[h]
\begin{center}
\begin{tabular}{|c|c|c|}
\hline
\multicolumn{1}{|c|} {\raisebox{-0.45ex}{Light}} &   
\multicolumn{2}{|c|} {Mass Limit} \\
\cline{2-3} 
&&\\[-1.6ex]
\multicolumn{1}{|c|} {\raisebox{0.45ex}{Squark}} & 
\multicolumn{1}{|c|} {$S/ \sqrt{B} = 5$} & 
\multicolumn{1}{|c|} {$S/ \sqrt{B} = 10$}     \\
\hline
$\tilde u_{L,R}$ & 590 & 510 \\
$\tilde d_{L,R}$ & 350 & 300 \\
$\tilde c_{L,R}$ & 265 & 225 \\
$\tilde s_{L,R}$ & 240 & 200 \\
$\tilde b_{L,R}$ & 155 & 125 \\
All       & 600 & 520 \\
\hline
\end{tabular}
\end{center}
   \caption{\em The mass limits that can be reached with $100 \pb^{-1}$ 
     integrated luminosity if only one flavour of squarks were light. 
     The last line represents the case when all the five flavours are 
     degenerate and corresponds to Fig.\protect\ref{fig:pT_distrib}. }
   \label{tab:5&10}
\end{table}
As Table~\ref{tab:5&10} illustrates, 
the bounds for the second generation squarks are also quite significant.
For example, even if only the $\tilde c_{L,R}$ were light,  
 $ S/\sqrt{B} > 5$ can still be obtained for 
$m_{\tilde c} \lsim 265 \gev$. 
Thus, if the proposed reanalysis of Tevatron data fails to 
produce any evidence for such a ($m_{\tilde g},\ m_{\tilde q}$) pair, 
it would effectively rule out, for example, the light gluino 
solutions~\cite{farrar_4jet,farr_new} to the 4-jet excess reported by 
ALEPH~\cite{aleph4j}. As for $\tilde b_{L,R}$, the relatively weak 
bounds of Table~\ref{tab:5&10} could be significantly improved if 
$b$-identification is used. The second potential drawback to our 
analysis is our deliberate ignoring of experimental efficiency factors. 
This, however, is unlikely to be a major factor. On the other hand, 
partial improvement might be possible if a more detailed fitting 
of the event distribution is attempted. These issues, though,  can be 
addressed only in a full simulation. 

With the ten-fold increase in luminosity that the main injector is 
expected to deliver, the $ S/\sqrt{B}$ ratios would essentially 
increase by a factor of $\sqrt{10}$ and thus the reach can be extended 
to even higher masses. For $\tilde u_{L,R}$, for example, the bounds would 
be close to 800 GeV. With the advent of LHC, the bounds for each 
squark flavour (and chirality) would tend to be well above 1~TeV, thus 
destroying all motivation for a light gluino.

To summarize, we have examined a particularly striking consequence 
of the light gluino scenario. While most of the mass range of interest
has already been ruled out, existing analyses have found 
it difficult to close the 
$m_{\tilde g} \lsim 1.5 \gev$ window.  We aver that a reanalysis of 
existing Tevatron data on monojets accompanied by large missing 
momentum can severely constrain this window too. 
As such a small mass for the gluino
results in a small, but nonnegligible, gluino content in the proton,
resonant production of squarks becomes possible. The subsequent decay 
of the squark into a quark and the LSP results in the signature described
above. We have performed a parton level simulation for both the 
signal and the SM backgrounds. For a properly chosen $p_T$ window, the 
signal is visible over 
the background for a considerably wide range of squark mass. The 
suggested reexamination of the existing Tevatron data can thus either 
establish this scenario or, in the case of a negative result, severely
constrain it. 

{\bf Acknowledgements:} I wish to thank Andreas Vogt and Reinhold R\"uckl
for sharing their codes for proton structure 
functions (ref.\cite{ruc_vog}). Thanks are due to Gautam Bhattacharyya and
D.P.~Roy for some very useful suggestions.

\end{document}